\documentclass[doublecol]{epl2} 
\usepackage{xcolor}
\usepackage{babel}
\usepackage{units}
\usepackage{textcomp}
\usepackage{mathrsfs}
\usepackage{amsmath}
\usepackage{amssymb}
\usepackage{graphicx}
\usepackage{wasysym}
\usepackage{esint}
\usepackage{nomencl}
\usepackage{bbold}
\usepackage{subcaption}
\usepackage{url}
\usepackage{gensymb}
\usepackage[normalem]{ulem}

\newcommand{\mc}[1]{\mathcal{{#1}}}

\title{Towards twisted, topological, and quantum  graphene plasmonics}
\shorttitle{Towards quantum and topological graphene plasmonics}

\author{A. Octávio Soares\inst{1,2} \and Nuno M. R. Peres \inst{2,3,4}}

\institute{  
\inst{1} Department of Physics and Astronomy, University of Porto, Rua do Campo Alegre, Porto, Portugal.\\
\inst{2} Department and Center of Physics (CF-UM-UP), University of Minho, Campus of Gualtar, Braga, Portugal\\
\inst{3} International Iberian Nanotechnology Laboratory (INL), Avenida José Mestre, Braga, Portugal \\
\inst{4} POLIMA---Center for Polariton-driven Light--Matter Interactions, University of Southern Denmark, Odense, Denmark
}

\abstract{
In this article, we analyze the quantum and topological properties of graphene-based plasmonic systems. We consider the following plasmonic materials: single-layer graphene, twisted bilayer graphene, and other graphene stackings, as well as the following architectures: graphene-based gratings, grids, chains of graphene disks, and the kagom\'e lattice.}

\begin{document}

\maketitle

\section{\label{sec:Plasmonics} A short introduction to plasmonics}

The field of plasmonics concerns the interaction between light and conduction electrons in materials. Historically, it is associated with the observation of Wood's anomalies in 1902, when sharp intensity drops were observed in the spectrum of visible light diffracted by metal gratings under specific conditions of incidence of $p$-polarized light. Several explanations and developments followed throughout the twentieth century, including the works of Lord Rayleigh and Fano ~\cite{enochPlasmonicsBasicsAdvanced2012}.

Plasmonics experienced renewed interest at the turn of the century, largely motivated by the observation of "Extraordinary optical transmission through subwavelength hole arrays" ~\cite{ebbesenExtraordinaryOpticalTransmission1998}. Theoretical descriptions of this phenomenon attributed the transmission peaks in periodically patterned silver films to the excitation of surface plasmon-polaritons (SPPs). These modes are electromagnetic (EM) surface waves that arise from the coupling between light and conduction electrons in metals, forming collective charge-density oscillations that propagate along the metal surface. In addition, SPP excitations exhibit exponentially decaying EM fields away from the interface, which results in subwavelength confinement and near-field intensities larger than that of the incident light, a property known as field enhancement ~\cite{maierPlasmonicsFundamentalsApplications2007}.

The modern field of plasmonics has developed alongside the broader area of nanophotonics ~\cite{maradudinModernPlasmonics2014}. The latter is primarily concerned with the study of optics at the nanometer scale. One of its main motivations is the ability to manipulate light at the nanoscale ~\cite{novotnyPrinciplesNanoOptics2012,hohenesterNanoQuantumOptics2020}, achieving confinement and control beyond the limits of conventional materials and geometries. This has been facilitated by advances in nanofabrication techniques ~\cite{yaoPlasmonicMetamaterials2014}. Within this context, plasmonics has assumed a leading role in the miniaturization of photonic structures. Due to their hybrid nature, combining light and electronic excitations, plasmonic systems have also contributed to bridging the gap between photonic devices and semiconductor technologies. This has led to applications beyond fundamental physics, including sensing, spectroscopy, photovoltaics, and radiation guiding, among others ~\cite{stockmanNanoplasmonicsPhysicsApplications2011}.

Plasmons can manifest in several forms depending on their physical nature, but they are generally classified into two categories within the nanophotonics framework. Non-propagating plasmons, typically arising at finite metal-dielectric interfaces such as nanoparticles, are known as localized surface plasmons (LSPs). Of greater relevance to this work are propagating plasmons supported by extended interfaces, known as SPPs ~\cite{goncalvesIntroductionGraphenePlasmonics2016}. Traditionally, the existence of SPPs required an interface between two materials with permittivities of opposite sign, such as a dielectric-metal interface. However, such excitations can also occur in two-dimensional conducting systems ~\cite{yoonPlasmonicsTwodimensionalConductors2014}, including semiconductor heterojunctions ~\cite{andressUltrasubwavelengthTwodimensionalPlasmonic2012} and graphene. In the following sections, we explore the rich research landscape associated with the latter.

\section{ A brief note on topology}
Topology in condensed matter physics
~\cite{araujoTopologyCondensedMatter2021}
traces its origins to the explanation of the quantum Hall effect in terms of the Chern number.
The recognition that many physical properties of a system—whether quantum or classical—can be understood in terms of topological invariants has opened new avenues of research in condensed matter physics, photonics, acoustics, electronics, and micromechanics. Across these fields, one can define physical quantities that are characterized by topological invariants, which classify distinct topological phases according to underlying symmetries and dimensionality, akin to order parameters.
Particularly noteworthy are the transport properties of electronic systems and the propagation of polaritons in platforms governed by strong light–matter interactions. For example, in finite-size two-dimensional photonic systems, electromagnetic fields can propagate along the system’s edges. These edge channels are robust against disorder and arise when the corresponding bulk topological invariant is quantized and nonzero. This result is dubbed the bulk-edge correspondence.
\section{\label{subsec:Graphene} The graphene case study}

Graphene is a two-dimensional material composed of a single layer of carbon atoms arranged in a honeycomb lattice. Since its first isolation ~\cite{novoselovElectricFieldEffect2004}, it has become a "wonder material" that has attracted significant attention in the condensed matter community.
Perhaps, the most notable characteristic is the presence of massless carriers associated with the Dirac cones in its band structure, which render pristine graphene a semi-metal ~\cite{netoElectronicPropertiesGraphene2009,peresColloquiumTransportProperties2010} with unusual physical properties.

Of equal importance in the optical and plasmonic study of graphene is its optical conductivity $\sigma_g(\omega)$, which encodes the response of the material and the interactions that occur in the presence of external electromagnetic fields. The peculiar band structure of graphene gives rise to remarkable optical properties, which have been extensively explored in plasmonics ~\cite{goncalvesIntroductionGraphenePlasmonics2016}.

Within linear response theory, the conductivity is typically computed using the Kubo formalism ~\cite{kuboStatisticalMechanicalTheoryIrreversible1957}, often combined with additional approximation schemes ~\cite{falkovskySpacetimeDispersionGraphene2007}. Alternative approaches can also be employed, such as the semiclassical Boltzmann transport equation ~\cite{chavesHydrodynamicModelApproach2017,narozhnyHydrodynamicApproachTwodimensional2022} or the polarizability tensor formalism ~\cite{hwangDielectricFunctionScreening2007,wunschDynamicalPolarizationGraphene2006}. The conductivity is commonly decomposed into two contributions: intraband and interband terms, corresponding to electronic transitions within the conduction band and between the conduction and valence bands, respectively. However, in doped graphene at low temperatures, in the regime $E_\mathrm{F}\gg k_\mathrm{B} T$, interband transitions contribute negligibly to the optical conductivity for incident frequencies $\omega < 2 E_\mathrm{F}/\hbar$. In practice, for frequencies in the terahertz (THz) to mid-infrared (mid-IR) range, even at room temperature, the conductivity is dominated by the intraband contribution, which takes a Drude-like form
$
\sigma_g(\omega)\approx \frac{i\sigma_0}{\hbar \pi}\frac{4E_F}{\omega +  i\gamma},
$
where $\sigma_0=e^2/4\hbar$ is the universal AC conductivity of graphene. In more refined models that include non-local effects, the conductivity depends on both the wave vector $\mathbf{q}$ and the frequency $\omega$. Since the imaginary part of the conductivity is positive, a graphene monolayer can support SPP excitations, known as graphene surface plasmon-polaritons (GSPPs), consisting of propagating oscillations of the carrier density. GSPPs were first observed in ~\cite{chenOpticalNanoimagingGatetunable2012,feiGatetuningGraphenePlasmons2012} and have since demonstrated remarkable advantages over traditional plasmonic systems based on noble metals such as gold and silver in the THz to mid-IR spectral range ~\cite{stauberPlasmonicsDiracSystems2014}. In this regime, GSPPs exhibit both longer propagation lengths and longer lifetimes compared to their metallic counterparts. Furthermore, graphene is highly tunable through various techniques, enabling the engineering of its plasmonic properties for applications in sensing and optoelectronics ~\cite{goncalvesIntroductionGraphenePlasmonics2016}.

As noted above, the Drude model for graphene conductivity can be extended to include non-local effects ~\cite{lundebergTuningQuantumNonlocal2017}, which become relevant when the size of nanostructures is sufficiently small that the electronic charge cannot be treated as locally homogeneous ~\cite{diasProbingNonlocalEffects2018}. In such cases, the optical conductivity must be treated as a function of both the frequency $\omega$ and the wave vector $\mathbf{q}$, thereby accounting for spatial dispersion. Several approaches have been developed to go beyond the local Drude approximation, including hydrodynamic models ~\cite{chavesHydrodynamicModelApproach2017,christensenClassicalQuantumPlasmonics2015}, the Lindhard formalism ~\cite{latyshevLongitudinalElectricConductivity2014,dasRelaxationtimeApproximationRPA1975,merminLindhardDielectricFunction1970}, and full Kubo calculations ~\cite{rodriguez-lopezElectricConductivityGraphene2025}. A recent review of non-local effects in photonic and plasmonic systems, including graphene, can be found in ~\cite{monticoneNonlocalityPhotonicMaterials2025}.

For plasmons at metal interfaces, non-local effects become important when the SPP wavelength becomes comparable to or smaller than the characteristic size of the nanostructure. In graphene, a commonly used criterion for the onset of non-locality is that the GSPP wave vector $q$ becomes large compared to the Fermi wave number $k_\mathrm{F}$, such that $qc\gg k_\mathrm{F} v_\mathrm{F}$, where $v_\mathrm{F}$ is the Fermi velocity ~\cite{chavesHydrodynamicModelApproach2017}. This effect is further enhanced when graphene is placed in close proximity to a metallic substrate, which increases screening and leads to the emergence of acoustic GSPPs with a linear dispersion relation $\omega \propto q$ ~\cite{epsteinFarfieldExcitationSingle2020}. These screened (or acoustic) plasmons possess significantly larger wavelengths than conventional GSPPs at the same frequency, thereby enhancing non-local effects.

Beyond single-layer graphene, considerable effort has been devoted to studying extended graphene-based systems. A natural extension is bilayer graphene, consisting of two graphene sheets stacked on top of each other ~\cite{mccannElectronicPropertiesBilayer2013}. Depending on the relative alignment of the layers, two primary stacking configurations arise, commonly referred to as AB (Bernal) and AA stacking ~\cite{netoElectronicPropertiesGraphene2009}. In contrast to monolayer graphene, AB-stacked bilayer graphene exhibits a parabolic dispersion that can be readily gapped and tuned by applying a bias voltage between the layers. Interlayer coupling also leads to the splitting of both valence and conduction bands into two branches ~\cite{minInitioTheoryGate2007}. Consequently, the optical response of bilayer graphene differs significantly from that of monolayer graphene. For example, a phonon resonance occurs at $\sim 0.2 \,\mathrm{eV}$, where strong enhancement of infrared activity arises when a plasmon resonance coincides with this frequency, leading to a characteristic Fano lineshape ~\cite{lowPlasmonsScreeningMonolayer2014}. It is worth noting that bilayer graphene should be distinguished from the related system of double-layer graphene, where two monolayers are separated by a dielectric spacer ~\cite{stauberPlasmonsNearfieldAmplification2012}. This latter system is closely related to acoustic GSPPs in graphene near metallic substrates, although it additionally supports an optical plasmon branch at higher frequencies.

Recently, there has also been growing interest in twisted bilayer graphene (TBG) ~\cite{lopesdossantosGrapheneBilayerTwist2007,andreiGrapheneBilayersTwist2020}, in which successive layers are rotated by a small relative angle, as well as in graphene multilayers, such as rhombohedral tetralayer graphene
~\cite{wirthExperimentalObservationABCB2022,choiSuperconductivityQuantizedAnomalous2025}
, which exhibits strongly correlated phases of matter ~\cite{liChargeFluctuationsPhonons2023}. In addition, trilayer graphene has been shown to exhibit stacking-dependent plasmon dispersion, as imaged using scattering-type scanning near-field optical microscopy (s-SNOM) ~\cite{luanImagingStackingDependentSurface2022}, a technique also widely used to image plasmons in other graphene-based systems ~\cite{alonso-gonzalezAcousticTerahertzGraphene2017,lundebergTuningQuantumNonlocal2017}.

Twisted bilayer graphene was the first graphene-based system in which strongly correlated phases of matter were experimentally observed ~\cite{caoUnconventionalSuperconductivityMagicangle2018,caoCorrelatedInsulatorBehaviour2018}. Early interest in such systems was motivated by the need to understand scanning tunneling microscopy (STM) observations, such as those reported in ~\cite{liObservationVanHove2010}, which revealed Moiré patterns—signatures of a relative twist between atomic layers—in bilayer graphene and graphene grown on $\mathrm{Si}\mathrm{C}$ substrates ~\cite{pongReviewOutlookAnomaly2005}. A major surge in research activity occurred following the discovery that TBG can host exotic phases of matter, including unconventional superconductivity ~\cite{caoUnconventionalSuperconductivityMagicangle2018}, when the relative twist angle is approximately $\theta=1.1^\circ$, giving rise to the field of twistronics ~\cite{andreiTwisted2DMaterials2024}. At this so-called magic angle, TBG exhibits flat electronic bands with bandwidths below $10\,\mathrm{meV}$ and a highly tunable carrier density, which suppresses kinetic energy and enhances the role of many-body interactions. Theoretical and experimental studies of the plasmonic response followed shortly thereafter, including the prediction and observation of plasmonic flat bands ~\cite{stauberQuasiFlatPlasmonicBands2016}, as well as chiral and slow plasmons ~\cite{huangObservationChiralSlow2022}, highlighting the high degree of tunability of the optical response in these systems.

Recently, TBG on talc heterostructures was experimentally investigated ~\cite{barbosaUltraconfinedPlasmonsReveal2025}.
Talc ~\cite{feresTwodimensionalTalcNatural2025} is a naturally occurring phyllosilicate and forms ultra-flat surfaces, very much like  hexagonal Boron Nitrite. When in contact with graphene induces p-type doping of the TBG. The TBG plasmons interact strongly with talc's phonon polaritons, leading to the formation of surface plasmon–phonon polariton modes.

\begin{figure*}[t]
    \centering
    \includegraphics[width=\textwidth]{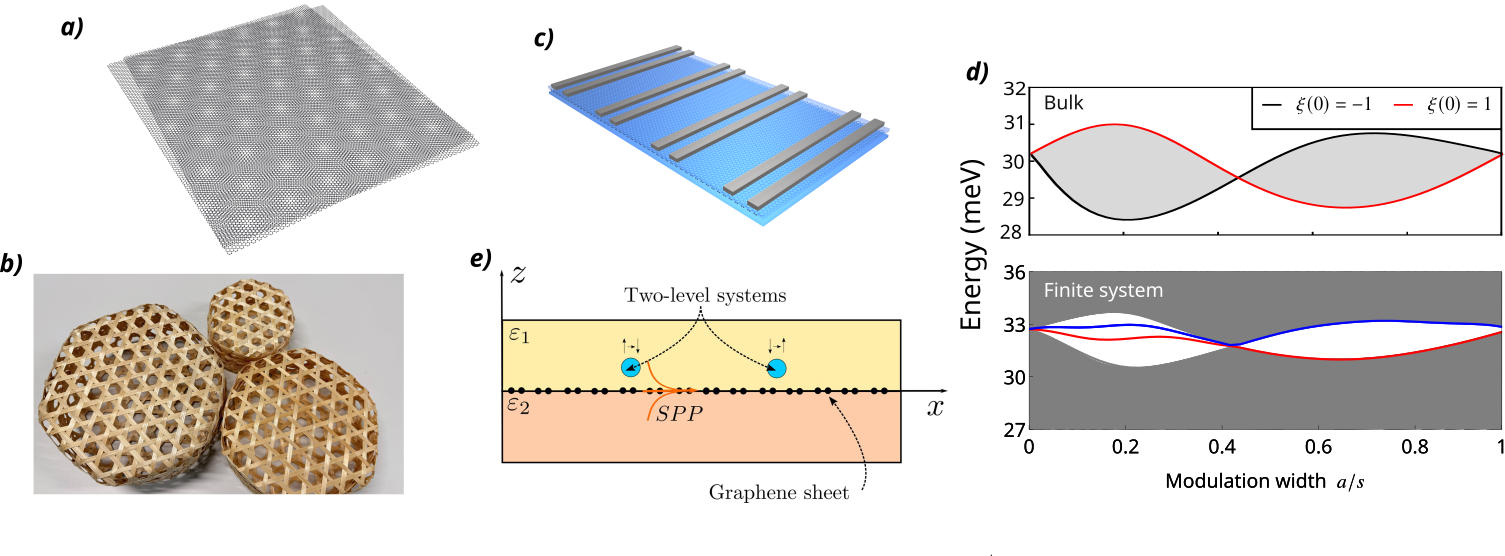}
        \caption{{\bf Systems discussed in the text.} a) Twisted bilayer graphene (TBG) with relative angle $\theta=4.1\degree$. b)   Basketweave kagomé lattice. c) Graphene plasmonic crystal realizing an SSH-like model via periodic metallic rods above a graphene sheet. d) Evolution of plasmonic energy levels with modulation width in a Kronig–Penney model. In the bulk, gap closing leads to parity exchange (band inversion). In a finite system, this corresponds to the emergence of mid-gap edge states, consistent with bulk–edge correspondence. e) Two-level systems coupled through graphene surface plasmon-polaritons (GSPPs).}
\end{figure*}
\section{\label{subsec:PlasmonicCrystals} Plasmonic crystals}
Over the past two decades, the field of photonics has experienced rapid development in the study of photonic crystals (PhCs). The construction of a PhC is conceptually simple, requiring only a spatial modulation of the optical properties of the medium through which light propagates. This is typically achieved by modulating the dielectric function or magnetic permeability, thereby imposing a periodic perturbation on Maxwell's equations ~\cite{joannopoulosPhotonicCrystalsMolding2011}, in direct analogy with a periodic potential in a crystalline solid. PhCs are regarded as one of the most effective approaches for controlling the flow of light, and one-dimensional (1D) PhCs are widely used as mirrors, known as distributed Bragg reflectors, that offer significant advantages over metallic alternatives. Similarly, PhCs can be engineered in two and three spatial dimensions, with bandgaps tailored to create materials that are transparent or opaque at selected frequencies. Furthermore, due to their high degree of tunability, PhCs provide a versatile platform for realizing condensed matter phenomena that are difficult to achieve in electronic systems, including photonic flat bands ~\cite{leykamPerspectivePhotonicFlatbands2018}, Anderson localization ~\cite{segevAndersonLocalizationLight2013}, and photonic analogues of quantum Hall states ~\cite{raghuAnalogsQuantumHalleffectEdge2008,ozawaTopologicalPhotonics2019}.

In a similar manner, structures supporting SPPs can be periodically modulated to form \emph{plasmonic crystals} (PlCs). One straightforward approach is to place a PhC in contact with an interface supporting SPP excitations, resulting in hybrid plasmon-photonic crystals ~\cite{xiongPhotonicCrystalGraphene2019}. Owing to its versatility, graphene provides an excellent platform for designing a wide variety of PlCs. One possible strategy is to arrange graphene nanoribbons in close proximity, allowing plasmons excited in one nanoribbon to couple to those in neighboring ribbons ~\cite{karimiPlasmonsGrapheneNanoribbons2017,nikitinSurfacePlasmonEnhanced2012,zhaoStrongPlasmonicCoupling2015}. Alternatively, one can modulate the surrounding environment of a graphene sheet, for example by placing metallic gratings beneath it ~\cite{bylinkinTightbindingTerahertzPlasmons2019,diasControllingSpoofPlasmons2017,jinTerahertzPlasmonicsFerroelectricgated2013,guoHybridGrapheneplasmonGratings2023}. It is also possible to introduce periodic modulation directly in the graphene itself, by varying its Fermi energy or carrier density ~\cite{azarGraphenePlasmonicCrystal2020,baeumerFerroelectricallyDrivenSpatial2015}, or by using sound waves. In the latter approach, a surface acoustic wave forms a diffraction grating that enables the excitation of the long-lived phonon-like branch of the hybrid graphene plasmon–phonon dispersion using infrared laser light ~\cite{schiefeleCouplingLightGraphene2013}.

To conclude this discussion, it is worth mentioning a recent research direction that focuses on the temporal modulation of spatially homogeneous photonic systems ~\cite{asgariTheoryApplicationsPhotonic2024}. This approach effectively creates one-dimensional crystals in which the quasiparticle momentum is conserved, while the frequency is not. Several analogues have been explored in plasmonic systems, including surface, bulk, and graphene plasmon-polaritons ~\cite{wilsonTemporalControlGraphene2018,kimGraphenePlasmonicTime2024,feinbergPlasmonicTimeCrystals2025}. By breaking time-translation symmetry, energy is no longer conserved, leading to new and potentially rich physical phenomena. Frequency bandgaps in conventional crystals are replaced by momentum bandgaps, which host states with purely imaginary momenta, corresponding to gain states with exponentially growing amplitudes that can enhance light-matter interactions ~\cite{lyubarovAmplifiedEmissionLasing2022}. However, achieving temporal modulations with sufficiently large amplitudes to observe such momentum bandgaps remains an open challenge, although several promising approaches are being explored ~\cite{wangExpandingMomentumBandgaps2025}.
\section{Topological graphene plasmons}
Using graphene as a platform for PlCs supporting propagating SPPs, it is possible to observe a wide range of topologically nontrivial phenomena. One approach consists of patterning a graphene sheet with a periodic structure and explicitly breaking time-reversal $\mc{T}$ symmetry by applying an external magnetic field, thereby inducing magnetoplasmon excitations and enabling the design of topological band structures. Although the prediction and observation of edge magnetoplasmons in conventional two-dimensional electron gases, as well as in graphene disks, date back over forty years, their connection to topological phases has only recently been established ~\cite{jinTopologicalMagnetoplasmon2016}. These systems typically support topologically protected edge states in the GHz regime. However, by introducing holes (or anti-dots) into the graphene sheet, or by employing graphene superlattices with hexagonal hole arrays arranged in a honeycomb geometry, topologically protected magnetoplasmons can be realized in the infrared spectral range ~\cite{panTopologicallyProtectedDirac2017,jinInfraredTopologicalPlasmons2017}.

An alternative strategy involves modulating the intrinsic properties of the graphene sheet. When such modulation is implemented along a single spatial direction, it becomes possible to realize plasmonic analogues of the Su-Schrieffer-Heeger (SSH) model. This approach has been explored, for example, by placing metallic rods on top of a graphene sheet ~\cite{mirandaTopologyOnedimensionalPlasmonic2024,tatianag.rappoportTopologicalGraphenePlasmons2021}, leading to topological behavior analogous to that of the SSH chain. Although the system cannot be mapped directly to the SSH model, similar behavior is also present when the Fermi level of graphene is varied periodically ~\cite{breyQuantumBandStructure2025,soaresScreenedTopologicalPlasmons2025}

\section{Quantum plasmons in graphene}
It is possible to observe purely quantum phenomena of SPPs, such as interference and decoherence, in various platforms including graphene ~\cite{piazzaSimultaneousObservationQuantization2015,lundebergTuningQuantumNonlocal2017}. Due to the hybrid nature of SPPs, the quantum properties can come from either their electronic or light components. In graphene structures, electronic quantum finite-size corrections to the optical conductivity only become relevant in very small structures, for instance in nanodisks smaller than 10 nm ~\cite{thongrattanasiriQuantumFiniteSizeEffects2012}. However, quantization of the EM fields driving the SPPs is necessary when performing nanoptics experiments ~\cite{tameQuantumPlasmonics2013}, \emph{i.e.} when the sources become coherent or are made up of single to few photons. In this regime, the fields can no longer be treated as classical ~\cite{tameSinglePhotonExcitationSurface2008}, and must be quantized following similar approaches to the canonical quantization of EM in vacuum. The quantization of plasmons in graphene has been carried out in a variety of setups ~\cite{hansonQuantumPlasmonicExcitation2015}, for instance in combination with an hydrodynamic model for the conductivity ~\cite{ferreiraQuantizationGraphenePlasmons2020,cardosoApplicationMadelungHydrodynamics2025} or in structured interfaces ~\cite{breyQuantumPlasmonsDouble2024,soaresScreenedTopologicalPlasmons2025}.

Dealing with SPPs in graphene as quantized quasi-particles has opened the possibility of combining these platforms in nanoptics experiments, for instance in studying their coupling to quantum emitters ~\cite{tormaStrongCouplingSurface2014} or two-level systems ~\cite{antaoTwolevelSystemsCoupled2021}. In addition, GSPPs are capable of storing and mediating quantum information ~\cite{sunGrapheneSourceEntangled2022}, with promising applications to quantum computing schemes ~\cite{alonsocalafellQuantumComputingGraphene2019,calajoNonlinearQuantumLogic2023}

\section{Open system effects on graphene plasmons}
Compared to conventional noble metal plasmonic systems, graphene is a compelling alternative platform because GSPPs exhibit longer propagation lengths and stronger field confinement than the former, particularly in the mid-IR spectral range. Nevertheless, even within this spectral window, multiple damping channels for GSPPs have been identified, including coupling to optical phonons of the substrates and scattering at edges ~\cite{yanDampingPathwaysMidinfrared2013}. At cryogenic temperatures and for sufficiently high doping levels, plasmon propagation is primarily limited by dielectric losses, with record propagation lengths of up to $\sim 50$ plasmon wavelengths have been reported ~\cite{woessnerHighlyConfinedLowloss2015,niFundamentalLimitsGraphene2018}. In all such cases, a comprehensive theoretical description of plasmon excitations must account for damping, losses, and potentially even optical gain. Within semiclassical approaches it is common to introduce a phenomenological damping parameter $\gamma$ in the conductivity, which effectively captures all loss mechanisms and is typically treated as a fitting parameter to experimental data ~\cite{goncalvesIntroductionGraphenePlasmonics2016}. However, this parameter should ideally be derived from first-principles calculations that explicitly incorporate the relevant plasmon decay channels. This problem has attracted some attention within the graphene plasmonics community ~\cite{sharmaOpticalConductivityDamping2024,liFirstprincipleCalculationsPlasmon2023,principiIntrinsicLifetimeDirac2013}.

When transitioning to a quantum-mechanical description, the non-conservation of energy is closely associated with non-Hermitian Hamiltonians $\mc{H}\neq \mc{H}^\dagger$.  This perspective underlies the framework of open quantum systems, and a variety of methods have been developed to analyze these systems, as reviewed in ~\cite{simObservablesNonHermitianSystems2025}. Importantly, non-Hermitian formulations are not limited to modeling losses and gains, but can also arise when certain degrees of freedom are traced out. This situation occurs, for example, in systems where one or more quantum emitters are coupled to a plasmonic structure, with plasmon-polariton modes mediating energy transfer and effective interactions between emitters ~\cite{cortesNonHermitianApproachQuantum2020,varguetNonhermitianHamiltonianDescription2019}. In such scenarios, the plasmonic modes act as an effective environment for the reduced emitter subsystem, and a complete description of the dynamics generally requires master-equation approaches, such as those based on the Lindblad formalism, rather than a purely Hermitian few-body Hamiltonian ~\cite{antaoTwolevelSystemsCoupled2021}. These techniques have been widely employed in quantum plasmonics to describe coherence transfer and the generation of entanglement between emitters mediated by plasmonic excitations ~\cite{gonzalez-tudelaEntanglementTwoQubits2011,leeRobusttolossEntanglementGeneration2013,sunGrapheneSourceEntangled2022}.

\section{Outlook}
Recently, the field of topological plasmonics in nanoparticle chains has attracted significant attention  ~\cite{mayerDirectObservationPlasmon2019}. A variety of condensed matter models have been realized using plasmonic nanostructures, including the SSH model ~\cite{yanNearFieldImagingTimeDomain2021} and the Aubry-André model ~\cite{yanNearfieldImagingSynthetic2025}. These systems typically operate in the visible spectral range, and extending their functionality to the mid-IR requires the use of alternative materials. In this context, one may envision chains of graphene micro-disks exhibiting topological properties in the mid-IR spectral range.
Another promising geometry is the kagom\'e lattice~\cite{syoziStatisticsKagomeLattice1951,mekataKagomeStoryBasketweave2003}. 
Finite clusters formed by repeating kagom\'e unit cells composed of graphene disks are expected to support both topological corner and edge states ~\cite{wangIntriguingKagomeTopological2025}, in addition to bulk modes. The potential use of such corner states in sensing applications is particularly appealing.
Furthermore, as material quality continues to improve, the implementation of quantum photonics experiments based on graphene plasmons (and other 2D materials and excitations) is expected to become increasingly feasible ~\cite{turunenQuantumPhotonicsLayered2022}.

\acknowledgments
A. O. Soares acknowledges FCT for a PhD scholarship under grant N$^{o.}$ 2025.00681.BD. N. M. R. Peres acknowledges his collaborators for valuable discussions on the topics of this article. The authors thank Katsunori Wakabayashi for the photography of the basketweave kagomé lattice.

\bibliographystyle{eplbib.bst}
\bibliography{Graphene_EPL_bib_file}

\end{document}